\documentclass[runningheads]{llncs}
\usepackage[T1]{fontenc}
\usepackage{graphicx}
\usepackage{amsmath,cite,url}
\usepackage{graphicx}
\usepackage{color}
\usepackage{cite}
\usepackage{amsmath,amssymb,amsfonts}
\usepackage{algorithmic}
\usepackage{graphicx}
\usepackage{textcomp}
\usepackage{multirow}
\usepackage{xcolor}
\usepackage{ragged2e}
\usepackage{array}
\usepackage{booktabs}
\usepackage{float}
\usepackage{tabularx}  
\begin{document}
\title{Training a Perceptual Model for Evaluating Auditory Similarity in Music Adversarial Attack}
\titlerunning{PAMT Perception Model}
%
\author{Yuxuan Liu\and
Rui Sang\and
Peihong Zhang\and
Zhixin Li\and
Shengchen Li} 
\authorrunning{Y. Liu et al.}
%
\institute{Xi'an Jiaotong-Liverpool University, Suzhou, 215123, China\\
\email{\{yuxuan.liu2204, rui.sang22, peihong.zhang20, zhixin.li22\}@student.xjtlu.edu.cn, shengchen.li@xjtlu.edu.cn}}
\maketitle              

\begin{abstract}
Music Information Retrieval (MIR) systems are highly vulnerable to adversarial attacks that are often imperceptible to humans, primarily due to a misalignment between model feature spaces and human auditory perception. Existing defenses and perceptual metrics frequently fail to adequately capture these auditory nuances, a limitation supported by our initial listening tests showing low correlation between common metrics and human judgments. To bridge this gap, we introduce Perceptually-Aligned MERT Transformer (PAMT), a novel framework for learning robust, perceptually-aligned music representations. Our core innovation lies in the psychoacoustically-conditioned sequential contrastive transformer, a lightweight projection head built atop a frozen MERT encoder. PAMT achieves a Spearman correlation coefficient of 0.65 with subjective scores, outperforming existing perceptual metrics. Our approach also achieves an average of 9.15\% improvement in robust accuracy on challenging MIR tasks, including Cover Song Identification and Music Genre Classification, under diverse perceptual adversarial attacks. This work pioneers architecturally-integrated psychoacoustic conditioning, yielding representations significantly more aligned with human perception and robust against music adversarial attacks.
\end{abstract}

\section{Introduction}
\label{sec:intro}

Music Information Retrieval (MIR) systems play an essential role in numerous applications, including music recommendation~\cite{musicrecommned1, musicrecommend2}, genre classification~\cite{genreicassp, genre2}, and cover song identification~\cite{du2021bytecover, hu2022wideresnetcover}. These systems rely on machine learning models to extract and interpret intricate musical features. However, recent studies have shown that MIR models are vulnerable to adversarial attacks, where small perturbations, imperceptible to human listeners, can cause significant prediction errors in these systems~\cite{saadatpanah2020adversarial, prinz2021end}. Such vulnerabilities not only threaten the reliability of MIR systems but also expose critical weaknesses in their robustness~\cite{saadatpanah2020adversarial}. Consequently, developing a perceptual model to quantify the auditory similarity between perturbed and original music is important for adversarial attacks and defenses.

Traditional adversarial attack and defense strategies often rely on mathematical norms (e.g., $L_p$ norms)~\cite{carlini2017towards, qin2019imperceptible} or signal-level metrics such as Signal-to-Noise Ratio (SNR) or Log-Spectral Distance (LSD)~\cite{gray1976distance} to evaluate the auditory similarity of adversarial examples. However, these metrics fail to account for the complexities of human auditory perception~\cite{duan2022perception}. For instance, due to phenomena such as auditory masking~\cite{wu2022catch, abdullah2021beyond}, the same perturbation magnitude may have varying perceptual effects depending on the spectral context. Such limitations highlight the disconnect between conventional perceptual metrics and human auditory perception.

Recent efforts have sought to incorporate human perception into adversarial attack evaluation. For example, psychoacoustic models have been integrated into adversarial attack design to better align perturbations with human auditory thresholds~\cite{schonherr2018adversarial}. However, these models often rely on fixed psychoacoustic parameters, making them less adaptable to diverse audio contexts and listener variations~\cite{chen2023imperceptible}. Another line of work leverages human auditory ratings to evaluate adversarial perturbations~\cite{duan2022perception}, but the use of non-differentiable alignment methods, such as Dynamic Time Warping, hinders integration with gradient-based deep learning frameworks~\cite{prinz2021end}. In music generation tasks~\cite{marafioti2020gacela, moliner2023solving, greshler2021catch}, researchers often utilize PEMO-Q~\cite{huber2006pemo} and Fréchet Audio Distance (FAD)~\cite{kilgour2019frechet} to measure the perceptual distance between generated music and the original track. However, there is currently no experimental evidence to confirm the effectiveness of these metrics in adversarial perturbation tasks.

To address these challenges, we conducted a human listening test to assess the efficacy of commonly used auditory metrics, including SNR, LSD, FAD, and PEMO-Q, in evaluating adversarial perturbations. Our results reveal that these metrics exhibit low to moderate correlation with human perceptual judgments, underscoring their limitations in accurately capturing the auditory similarity of adversarial examples. As shown in Figure~\ref{fig:metric_correlation_updated}, pre-trained music representation models yield a Fréchet Audio Distance (FAD) more closely aligned with human judgments than traditional signal-based metrics. Inspired by recent advances in music representation learning, particularly the MERT (Acoustic Music Understanding Model with Large-Scale Self-supervised Training~\cite{li2023mert}, we hypothesize that such pre-trained representations can be further refined to better align with human perception of adversarial perturbations.

However, standard music representation models like MERT are not explicitly designed to be robust against imperceptible adversarial perturbations. To address this limitation, we propose Perceptually-Aligned MERT Transformer (PAMT), a novel approach that leverages MERT's rich music representations and projects them into a perceptually-aligned feature space. Our method employs a Transformer-based projection head that is conditioned on psychoacoustic perturbation parameters through Feature-wise Linear Modulation (FiLM)~\cite{riou2025FiLM} and trained with a sequential contrastive objective. This design allows the model to learn representations that remain consistent for perceptually similar inputs while differentiating between perceptually distinct samples, even when the mathematical differences are subtle. By explicitly conditioning on psychoacoustic perturbation characteristics, our model becomes sensitive to the context-dependent nature of human auditory perception. This context sensitivity can help detect subtle perturbations in adversarial examples.

\section{Human Listening Test and Metric Evaluation}
\label{sec:listening_test_and_metrics}

To motivate our approach, we first conducted a human listening study to quantify the perceptual similarity between original and adversarially perturbed music. We then evaluated how well existing objective auditory metrics correlate with these human judgments, thereby demonstrating the need for a more perceptually-aligned model.

\subsection{Data Preparation for Subjective Evaluation}
\label{ssec:data_prep_subjective}

We created a dataset of reference and perturbed audio pairs from 1,000 unique tracks sampled across the FMA~\cite{defferrard2017fma}, GTZAN~\cite{tzanetakis2002musical}, and MTG-Jamendo~\cite{bogdanov2019mtg} repositories. All tracks were resampled to 16kHz and segmented into 10-second clips.

We applied six distinct adversarial perturbation types to the reference clips, as detailed in Table~\ref{tab:audio_distortions}, reflecting common attack methodologies from the literature~\cite{carlini2017towards, wu2022catch, duan2022perception}. By varying the parameters for each type, we generated 18,000 unique reference-perturbed pairs for subjective evaluation.

\begin{table}[t]
\caption{The six adversarial perturbation types simulate common adversarial attacks~\cite{carlini2017towards, wu2022catch, duan2022perception}. Perturbations are applied with varying parameter ranges to capture different attack strengths and characteristics.}
\label{tab:audio_distortions}
\centering
\renewcommand{\arraystretch}{1.2} 
\begin{tabular}{@{}>{\centering\arraybackslash}m{3.6cm} >{\RaggedRight\arraybackslash}m{8.6cm}@{}} 
\toprule
\textbf{Category} & \textbf{Perturbation Type (Parameter Range)} \\
\midrule
\vspace{2ex} 
\multirow{2}{*}{\textbf{$L_p$-based}} &
\begin{minipage}[t]{\linewidth} 
\textbf{$L_2$ noise:} Gaussian noise with $\|\delta\|_2 \leq \epsilon$, $\epsilon \in [0.01, 1.0] \times \text{RMS}_{\text{signal}}$. \\
\textbf{$L_\infty$ noise:} Uniform noise with $\|\delta\|_\infty \leq \eta$, $\eta \in [0.001, 0.01] \times \text{max}(|\text{signal}|)$.
\end{minipage} \\
\cmidrule{1-2}
\multirow{1}{*}{\textbf{Psychoacoustic-based}} &
\textbf{Bark-band noise:} Additive noise in a randomly selected Bark band, scaled by a factor of $[0.1, 0.5]$ relative to the original band energy. \\
\cmidrule{1-2}
\multirow{2}{*}{\textbf{Semantic-based}} &
\begin{minipage}[t]{\linewidth} 
\textbf{Pitch shift:} $n \in \mathcal{U}(-5, 5)$ semitones. \\
\textbf{Speed change:} $s \in \mathcal{U}(0.80, 1.20)$ factor.
\end{minipage} \\
\cmidrule{1-2}
\multirow{1}{*}{\textbf{Dynamics-based}} &
\textbf{Dynamic range compression:} Threshold $[-30, -10]$ dBFS, ratio $[2:1, 8:1]$. \\
\bottomrule
\end{tabular}
\end{table}

\subsection{Subjective Listening Test Design}
\label{ssec:subjective_test_design}

We conducted two listening experiments with 200 volunteers (50 with music backgrounds), who used their own high-quality headphones in quiet environments.

\subsubsection{MOS-style Perceptual Similarity Rating}
\label{sssec:mos_rating}
In our first task, participants rated the perceptual similarity of reference-perturbed pairs on a 5-point scale (1: identical, 5: dissimilar). Each of the 18,000 pairs was rated by at least five volunteers, yielding our "raw scores" dataset.

\subsubsection{Two-Alternative Forced Choice (2AFC) Test}
\label{sssec:2afc_test}
To mitigate rating biases~\cite{nayem2023attention, zhang2018unreasonable}, we also used a 2AFC test. For each reference audio $S_{\text{ref}}$, participants were presented with two of its perturbed versions, $S_{pi}$ and $S_{pj}$, and chose which sounded more similar to $S_{\text{ref}}$. For each of the 3,000 reference tracks, all 15 unique pairs of its perturbations were compared ($\binom{6}{2} \times 3000 = 45,000$ trials). We derived a "2AFC score" for each perturbed sample by counting how many times it was chosen as more similar, forming our "2AFC scores" dataset.

\subsection{Evaluation of Existing Objective Auditory Metrics}
\label{ssec:objective_metric_eval}

We then evaluated a suite of objective metrics against our human perceptual data. The metrics included traditional, psychoacoustic, and deep learning-based models:
\begin{itemize}
    \item \textbf{Signal-to-Noise Ratio (SNR):} Ratio of signal to perturbation power.
    \item \textbf{Log-Spectral Distance (LSD)~\cite{gray1976distance}:} MSE between log-magnitude spectra.
    \item \textbf{PEMO-Q~\cite{huber2006pemo}:} Psychoacoustic audio quality model.
    \item \textbf{Fréchet Audio Distance (FAD)~\cite{kilgour2019frechet}:} Using embeddings from VGGish~\cite{hershey2017cnn}, PANNS~\cite{kong2020panns}, and MERT-v0~\cite{li2023mert}.
    \item \textbf{Speech Quality/Perception Models:} CDPAM~\cite{manocha2021cdpam}, NOMAD~\cite{ragano2024nomad}, and SESQA~\cite{serra2021sesqa} for cross-domain evaluation.
\end{itemize}

\begin{figure}[t]
\centering
\includegraphics[width=\linewidth]{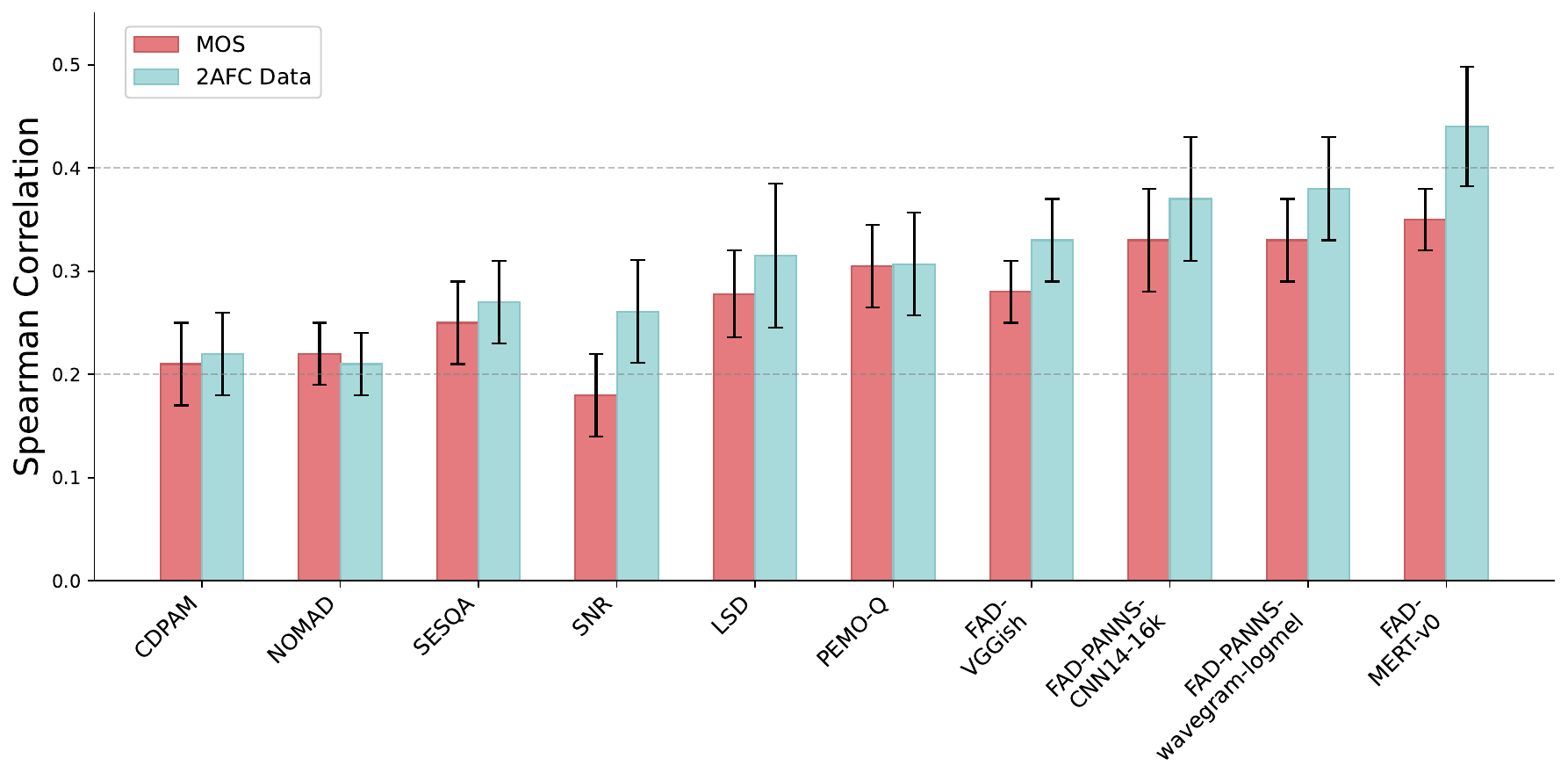} 
\caption{Spearman Correlation Coefficients ($\rho$) between objective auditory metrics and human perceptual similarity ratings (mean $\pm$ std. dev.). Correlations are shown for both raw MOS-style scores and 2AFC-derived scores. Objective metrics include SNR, LSD, PEMO-Q, FAD with various embeddings (VGGish~\cite{hershey2017cnn}, PANNS-CNN14-16k~\cite{kong2020panns}, PANNS-Wavegram-Logmel~\cite{kong2020panns}, MERT-v0~\cite{li2023mert}), and speech-domain models (CDPAM~\cite{manocha2021cdpam}, NOMAD~\cite{ragano2024nomad}, SESQA~\cite{serra2021sesqa}).}
\label{fig:metric_correlation_updated}
\end{figure}

We quantified the agreement by calculating the Spearman rank correlation ($\rho$) between each metric and our human scores (both raw and 2AFC). The results are presented in Figure~\ref{fig:metric_correlation_updated}.

As shown in Figure~\ref{fig:metric_correlation_updated}, existing metrics exhibited low to moderate correlation with human judgments. For raw MOS-style scores, correlations ranged from $\rho=0.18$ (SNR) to $\rho=0.35$ (FAD with MERT-v0). The more robust 2AFC scores yielded slightly better results, with FAD-MERT-v0 achieving the highest correlation ($\rho=0.44$). Speech-domain models showed similarly limited performance.

These results highlight a significant gap: even the best-performing metric, FAD with MERT-v0 embeddings, achieves only a modest correlation with robust human perceptual data. This limitation of existing metrics motivates our work on PAMT, a model that learns representations directly aligned with human auditory perception for enhanced robustness.

\section{PAMT: Learning Perceptually-Aligned Representations via Conditioned Sequential Contrastive Learning}
\label{sec:proposed_pamt}

The established limitations of existing objective metrics in capturing human perceptual similarity under adversarial conditions (Section~\ref{sec:listening_test_and_metrics}) necessitate novel representation learning techniques that explicitly embed principles of auditory perception. To this end, we introduce the Perceptually-Aligned MERT Transformer (PAMT) framework. PAMT transforms features from the general-purpose music understanding model, MERT~\cite{li2023mert}, into a new latent space that is demonstrably robust to imperceptible perturbations and closely aligned with human auditory perception. The core of the PAMT framework is our novel projection head: the Psychoacoustically-Conditioned Sequential Contrastive Transformer (PCSCT).

\subsection{PAMT Framework Overview}
\label{ssec:pamt_overview}

\begin{figure}[t]
\centering
\includegraphics[width=1\linewidth]{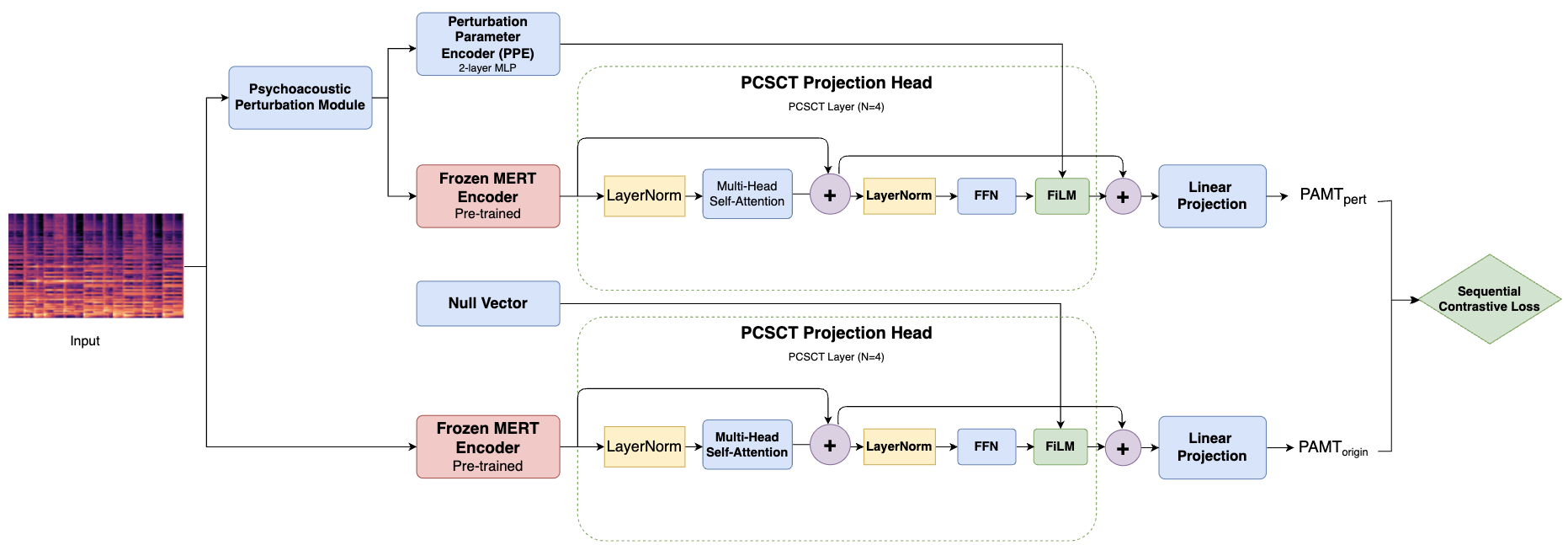} 
\caption{\textbf{Architectural Overview of the Proposed Perceptually-Aligned MERT Transformer (PAMT) Framework.} Audio is processed by a frozen MERT encoder. For training, a psychoacoustic perturbation module generates perturbed audio and its parameters. A Perturbation Parameter Encoder (PPE) creates a conditioning vector $c_{\text{perturb}}$ from these parameters. The core PCSCT projection head (a Transformer) processes MERT's sequential embeddings, conditioned by $c_{\text{perturb}}$ via FiLM layers, to produce robust, perceptually-aligned sequential embeddings $Z_{\text{PAMT}}$.}
\label{fig:pamt_architecture_overview}
\end{figure}

The PAMT framework, illustrated in Figure~\ref{fig:pamt_architecture_overview}, uses a frozen MERT encoder for feature extraction. These features are then refined by our novel PCSCT projection head, which is trained via a sequential contrastive loss and conditioned on psychoacoustic perturbation parameters.

The key components are:
\begin{enumerate}
    \item \textbf{Frozen MERT Encoder:} A pre-trained, frozen MERT-v0 model~\cite{li2023mert} extracts a sequence of 768-dimensional embeddings, $E_{\text{MERT}} \in \mathbb{R}^{T \times 768}$, from the input audio (resampled to 24kHz).
    \item \textbf{Psychoacoustic Perturbation Module:} During training, this module creates a perturbed audio version $A_{\text{pert}}$ from a reference $A_{\text{orig}}$ using imperceptible alterations (see Table~\ref{tab:audio_distortions}) and outputs the perturbation parameters $P_{\text{params}}$.
    \item \textbf{Perturbation Parameter Encoder (PPE):} A 2-layer MLP encodes $P_{\text{params}}$ into a compact 64-dimensional conditioning vector $c_{\text{perturb}}$.
    \item \textbf{PCSCT Projection Head:} This core component (detailed in Sec.~\ref{ssec:pcsct_head}) is a Transformer that processes the MERT sequence $E_{\text{MERT}}$, conditioned on $c_{\text{perturb}}$, to produce robust 128-dim embeddings $Z_{\text{PAMT}}$.
    \item \textbf{Sequential Contrastive Learning:} The PCSCT is trained to minimize the distance between representations of original and perturbed audio pairs ($Z_{\text{PAMT}}^{\text{orig}}$, $Z_{\text{PAMT}}^{\text{pert}}$) while maximizing their distance to other samples in the batch.
\end{enumerate}

\subsection{Psychoacoustically-Conditioned Sequential Contrastive Transformer (PCSCT)}
\label{ssec:pcsct_head}

The PCSCT projection head transforms MERT features into a new space where imperceptible psychoacoustic variations are minimized. This is achieved by explicitly conditioning the transformation on the perturbation's characteristics.

\subsubsection{Architecture}
PCSCT is a 4-layer Transformer encoder with 4 attention heads, a hidden dimension of 256, and a position-wise FFN with an inner dimension of 1024. We use GELU activations and a Pre-LN configuration~\cite{riou2025FiLM}. It takes the 768-dim $E_{\text{MERT}}$ sequence as input and, after a final linear projection, outputs a 128-dim sequence $Z_{\text{PAMT}}$.

\subsubsection{Psychoacoustic Conditioning with FiLM}
The key innovation is conditioning via Feature-wise Linear Modulation (FiLM) layers~\cite{riou2025FiLM}. The 64-dim conditioning vector $c_{\text{perturb}}$ from the PPE is projected to generate modulation parameters $(\gamma_l, \beta_l) \in \mathbb{R}^{256}$ for each Transformer layer $l$. These modulate the output $h_l$ of the FFN sub-layer:
\begin{equation}
    \text{FiLM}(h_l, c_{\text{perturb}}) = \gamma_l(c_{\text{perturb}}) \odot h_l + \beta_l(c_{\text{perturb}})
\end{equation}
where $\odot$ is element-wise multiplication. This allows PCSCT to adapt its feature transformation based on the specific perturbation profile to which it must learn invariance.

\subsubsection{Sequential Contrastive Learning Objective}
PCSCT is trained with an InfoNCE loss. For each pair of reference $E_{\text{MERT}}^{\text{orig}}$ and perturbed $E_{\text{MERT}}^{\text{pert}}$ embeddings, we generate conditioned outputs $Z_{\text{PAMT}}^{\text{orig}}$ and $Z_{\text{PAMT}}^{\text{pert}}$. The loss is:
\begin{equation}
\begin{aligned}
\mathcal{L}_{\text{PCSCT}} = - \mathbb{E} \Bigg[ \log \frac{ \exp\left(\frac{\text{sim}(Z_{\text{PAMT}}^{\text{orig}}, Z_{\text{PAMT}}^{\text{pert}})}{\tau}\right) }{
    \begin{aligned}[t] 
      & \sum_{\substack{k \in \text{Batch} \\ k \neq \text{orig}}} \exp\left(\frac{\text{sim}(Z_{\text{PAMT}}^{\text{orig}}, Z_{\text{PAMT}}^{k})}{\tau}\right) \\
      & \qquad + \exp\left(\frac{\text{sim}(Z_{\text{PAMT}}^{\text{orig}}, Z_{\text{PAMT}}^{\text{pert}})}{\tau}\right)
    \end{aligned}
  } \Bigg]
\end{aligned}
\end{equation}
The similarity function, $\text{sim}(U, V)$, is the cosine similarity of the time-pooled mean vectors of sequences $U$ and $V$. The temperature $\tau$ is set to 0.1.

\subsection{Guiding Adversarial Defense}
\label{ssec:pamt_defense}
Once trained, the PAMT embeddings $Z_{\text{PAMT}}$ (specifically, the mean-pooled vectors $\bar{z}_{\text{PAMT}}$) provide a robust feature space for downstream MIR tasks. For adversarial defense, we use adversarial training within this space. The perceptual distance $d_{\text{PAMT}}(A', A)$ is defined as the L2 norm between the PAMT embeddings of two audio clips $A'$ and $A$:
\begin{equation}
\label{eq:dpamt_definition}
d_{\text{PAMT}}(A', A) = \left\| \text{mean\_pool}(Z_{\text{PAMT}}(A')) - \text{mean\_pool}(Z_{\text{PAMT}}(A)) \right\|_2.
\end{equation}
Adversarial examples $A'$ are crafted to maximize the task loss while being constrained by $d_{\text{PAMT}}(A', A) \le \varepsilon$. The downstream model $f_{\theta}$ is trained by solving:
\begin{equation}
  \label{eq:pat_pamt}
  \min_{\theta} \;\; \mathbb{E}_{(A,y)\sim \mathcal{D}} \Bigl[\,
    \max_{d_{\text{PAMT}}(A',A)\,\le\, \varepsilon} \;\; \mathcal{L}_{\text{CE}}\bigl(f_{\theta}(\text{mean\_pool}(Z_{\text{PAMT}}(A'))),\,y\bigr)
  \Bigr],
\end{equation}
where $\mathcal{L}_{\text{CE}}$ is the cross-entropy loss. This strategy leverages the perceptual robustness of $Z_{\text{PAMT}}$ for stronger defense.
\section{Experiments}
\label{sec:experiments}

This section validates our Perceptually-Aligned MERT Transformer (PAMT), evaluating its ability to learn perceptually-aligned representations and its effectiveness in enhancing adversarial defense.

\subsection{Datasets and Evaluation Protocol}
\label{ssec:datasets_protocol_exps}

We use the human listening test dataset from Section~\ref{sec:listening_test_and_metrics}, annotated with both MOS and the more robust 2AFC perceptual similarity scores. For all evaluations, we use an 80/20 train/test split. Our primary evaluation metric is the Spearman rank correlation ($\rho$) between a model's predicted similarity and the human scores on the test set.

\subsection{Methods Compared}
\label{ssec:methods_compared_exps}

We compare PAMT against two main categories of methods:
\begin{enumerate}
    \item \textbf{Objective Auditory Metrics:} This includes the full suite of signal-based, psychoacoustic, and FAD-based metrics previously evaluated in Section~\ref{ssec:objective_metric_eval}, which serve as established benchmarks.
    \item \textbf{Learning-based Representation Models:} Our primary comparison is against a strong baseline and our proposed model.
\end{enumerate}
\begin{itemize}
    \item \textbf{MERT+MLP-Contrastive:} A strong baseline using the same frozen MERT-v0 encoder and psychoacoustic augmentations as PAMT, but with a simpler MLP projection head trained with standard contrastive loss. This isolates the benefits of our PCSCT architecture.
    \item \textbf{PAMT (PCSCT - Ours):} Our proposed framework, featuring the psychoacoustically conditioned, sequence-aware PCSCT projection head trained with a sequential contrastive loss.
\end{itemize}

\subsection{Training Details for PAMT (PCSCT)}
\label{ssec:training_details_pamt}

We train the PCSCT head using the AdamW optimizer (LR = $1 \times 10^{-4}$, weight decay = $1 \times 10^{-5}$) with a batch size of 32. We use a cosine annealing schedule with a 10\% warm-up and train for up to 100 epochs, with early stopping based on the validation set Spearman correlation (patience=10). The contrastive loss temperature $\tau$ is 0.1. The PPE is a 2-layer ReLU MLP producing a 64-dim vector. Experiments were run on NVIDIA A100 GPUs using PyTorch.

\subsection{Correlation with Human Perception}
\label{ssec:correlation_results_exps}

Table~\ref{tab:results_correlation_pamt} presents the Spearman correlation ($\rho$) and a classification-based $F_1$ score against the 2AFC human judgments. The $F_1$ score evaluates the model's utility in distinguishing noticeable from imperceptible perturbations.

\begin{table}[t!]
\centering
\caption{Evaluation of Perceptual Alignment on the Test Set. Spearman correlation ($\rho$) and $F_1$ score (\%) are reported against 2AFC human similarity judgments. Higher values are better.}
\label{tab:results_correlation_pamt}
\small
\begin{tabular}{p{0.62\columnwidth} >{\raggedleft\arraybackslash}p{0.15\columnwidth} >{\raggedleft\arraybackslash}p{0.1\columnwidth}}
\toprule
\textbf{Method} & \textbf{Spearman ($\rho$)} & \textbf{$F_1$(\%)} \\
\midrule
\textit{Traditional \& Speech Metrics} & & \\
\quad SNR & 0.261 & 56.1 \\
\quad LSD & 0.315 & 57.0 \\
\quad PEMO-Q & 0.307 & 58.1 \\
\quad CDPAM & 0.22 & 53.5 \\
\quad NOMAD & 0.21 & 53.1 \\
\quad SESQA & 0.27 & 55.2 \\
\midrule
\textit{FAD with Pre-trained Embeddings} & & \\
\quad FAD (VGGish) & 0.33 & 57.9 \\
\quad FAD (PANNS-CNN14-16k) & 0.37 & 59.5 \\
\quad FAD (PANNS-Wavegram-Logmel) & 0.38 & 60.3 \\
\quad FAD (MERT-v0 raw) & 0.44 & 63.1 \\
\midrule
\textit{MERT with Projection Heads} & & \\
\quad MERT + MLP (Std. Contrastive) & 0.55 & 68.5 \\ 
\quad \textbf{PAMT (PCSCT - Ours)} & \textbf{0.65} & \textbf{81.2} \\ 
\bottomrule
\end{tabular}
\end{table}

The results clearly show PAMT's superiority. It achieves a Spearman correlation of $\rho=0.65$, substantially outperforming all other methods, including FAD with raw MERT embeddings ($\rho=0.44$) and the strong MERT+MLP contrastive baseline ($\rho=0.55$). The high $F_1$ score of $81.2\%$ further confirms its effectiveness. This significant improvement demonstrates that our psychoacoustic conditioning and sequence-aware training are crucial for learning representations that genuinely align with human auditory perception.

\subsection{Effectiveness in Adversarial Defense}
\label{ssec:defense_results_exps}

We test PAMT's utility for adversarial defense on Cover Song Identification (CSI) and Music Genre Classification (MGC). We compare standard training (\textit{No Defense}), adversarial training in the input audio space (\textit{Standard AT}), and AT in the embedding spaces of our \textit{MERT+MLP} baseline and our proposed \textit{PAMT}. Robustness is measured by the worst-case performance ("Union Robust Acc.") against a set of strong perceptual attacks.

\begin{table}[t!]
\centering
\caption{Adversarial defense performance on CSI (mAP) and MGC (Accuracy). "Union Robust Acc." reflects the worst-case performance under a diverse set of perceptual adversarial attacks.}
\label{tab:mir_adv_results_pamt}
\small
\resizebox{\columnwidth}{!}{%
\begin{tabular}{lcc|cc}
\toprule
& \multicolumn{2}{c}{\textbf{CoverHunter (mAP)}}
& \multicolumn{2}{c}{\textbf{MGC ResNet (Acc)}} \\
\textbf{Method} & Clean & Union Robust Acc. & Clean & Union Robust Acc. \\
\midrule
No Defense & 0.845 & 0.021 & 0.828 & 0.035 \\
Standard AT ($L_p$) & 0.820 & 0.315 & 0.805 & 0.330 \\
MERT + MLP AT & 0.831 & 0.415 & 0.812 & 0.402 \\
\textbf{PAMT AT (Ours)} & \textbf{0.835} & \textbf{0.535} & \textbf{0.818} & \textbf{0.465} \\
\bottomrule
\end{tabular}
}
\end{table}

As shown in Table~\ref{tab:mir_adv_results_pamt}, undefended models collapse under perceptual attacks. Critically, adversarial training in our PAMT embedding space yields the highest robust accuracy in both CSI (0.535 mAP) and MGC (0.465 Acc), significantly outperforming standard AT and AT in the baseline embedding space, all while maintaining high performance on clean data. This confirms that the superior perceptual alignment of PAMT embeddings translates directly into more effective and practical adversarial defenses.
\section{Conclusions}
\label{sec:conclusions}

This paper introduced the Perceptually-Aligned MERT Transformer (PAMT) framework, centered on our novel Psychoacoustically-Conditioned Sequential Contrastive Transformer (PCSCT) projection head. Addressing the significant limitations of existing metrics in capturing human perception of adversarial audio perturbations—a gap validated by our listening tests—PAMT learns robust, perceptually-aligned music representations from frozen MERT encoders. The PCSCT's core innovation lies in its Transformer architecture, explicitly conditioned on psychoacoustic perturbation parameters via FiLM, and trained with a sequential contrastive objective to achieve invariance to imperceptible distortions. Our experiments demonstrate PAMT's superior alignment with human perceptual judgments (Spearman $\rho=0.65$) and its ability to substantially enhance adversarial robustness in downstream MIR tasks, such as Cover Song Identification and Music Genre Classification (achieving up to 0.535 mAP / 0.465 Accuracy under diverse attacks). This work pioneers an architecturally-integrated psychoacoustic conditioning approach, significantly advancing the development of MIR systems that are both more robust and more aligned with human auditory experience.
%
%
\bibliographystyle{splncs04}
\bibliography{CMMR2025_LaTeX2e_Proceedings_Templates/mybibliography}

\begin{thebibliography}{10}
\providecommand{\url}[1]{\texttt{#1}}
\providecommand{\urlprefix}{URL }
\providecommand{\doi}[1]{https://doi.org/#1}

\bibitem{abdullah2021beyond}
Abdullah, H., Rahman, M.S., Peeters, C., Gibson, C., Garcia, W., Bindschaedler, V., Shrimpton, T., Traynor, P.: Beyond $ l\_p $ clipping: Equalization based psychoacoustic attacks against asrs. In: Asian Conference on Machine Learning. pp. 672--688. PMLR (2021)

\bibitem{bogdanov2019mtg}
Bogdanov, D., Won, M., Tovstogan, P., Porter, A., Serra, X.: The mtg-jamendo dataset for automatic music tagging. In: Machine Learning for Music Discovery Workshop, International Conference on Machine Learning (ICML). Long Beach, CA, United States (2019), \url{http://hdl.handle.net/10230/42015}, iCML 2019

\bibitem{carlini2017towards}
Carlini, N., Wagner, D.: Towards evaluating the robustness of neural networks. In: 2017 IEEE Symposium on Security and Privacy (SP). pp. 39--57. IEEE (2017)

\bibitem{chen2023imperceptible}
Chen, L., Wang, R., Dong, L., Yan, D.: Imperceptible adversarial audio steganography based on psychoacoustic model. Multimedia Tools and Applications  \textbf{82}(17),  26451--26463 (2023)

\bibitem{defferrard2017fma}
Defferrard, M., Benzi, K., Vandergheynst, P., Bresson, X.: {FMA}: A dataset for music analysis. In: 18th International Society for Music Information Retrieval Conference (2017)

\bibitem{du2021bytecover}
Du, X., Yu, Z., Zhu, B., Chen, X., Ma, Z.: Bytecover: Cover song identification via multi-loss training. In: ICASSP 2021-2021 IEEE International Conference on Acoustics, Speech and Signal Processing (ICASSP). pp. 551--555. IEEE (2021)

\bibitem{duan2022perception}
Duan, R., Qu, Z., Zhao, S., Ding, L., Liu, Y., Lu, Z.: Perception-aware attack: Creating adversarial music via reverse-engineering human perception. In: Proceedings of the 2022 ACM SIGSAC Conference on Computer and Communications Security. pp. 905--919 (2022)

\bibitem{genre2}
Ghildiyal, A., Singh, K., Sharma, S.: Music genre classification using machine learning. In: 2020 4th International Conference on Electronics, Communication and Aerospace Technology (ICECA). pp. 1368--1372. IEEE (2020)

\bibitem{gray1976distance}
Gray, A., Markel, J.: Distance measures for speech processing. IEEE Transactions on Acoustics, Speech, and Signal Processing  \textbf{24}(5),  380--391 (1976)

\bibitem{greshler2021catch}
Greshler, G., Shaham, T., Michaeli, T.: Catch-a-waveform: Learning to generate audio from a single short example. Advances in Neural Information Processing Systems  \textbf{34},  20916--20928 (2021)

\bibitem{hershey2017cnn}
Hershey, S., Chaudhuri, S., Ellis, D.P., Gemmeke, J.F., Jansen, A., Moore, R.C., Plakal, M., Platt, D., Saurous, R.A., Seybold, B.: {CNN architectures for large-scale audio classification}. In: Proceedings of the IEEE International Conference on Acoustics, Speech and Signal Processing (ICASSP) (2017)

\bibitem{hu2022wideresnetcover}
Hu, S., Zhang, B., Lu, J., Jiang, Y., Wang, W., Kong, L., Zhao, W., Jiang, T.: Wideresnet with joint representation learning and data augmentation for cover song identification. In: Interspeech. pp. 4187--4191 (2022)

\bibitem{huber2006pemo}
Huber, R., Kollmeier, B.: {PEMO-Q—A} new method for objective audio quality assessment using a model of auditory perception. IEEE Transactions on Audio, Speech, and Language Processing  \textbf{14}(6),  1902--1911 (2006)

\bibitem{genreicassp}
Hung, Y.N., Yang, C.H.H., Chen, P.Y., Lerch, A.: Low-resource music genre classification with cross-modal neural model reprogramming. In: ICASSP 2023-2023 IEEE International Conference on Acoustics, Speech and Signal Processing (ICASSP). pp.~1--5. IEEE (2023)

\bibitem{kilgour2019frechet}
Kilgour, K., Zuluaga, M., Roblek, D., Sharifi, M.: Fréchet {A}udio {D}istance: A reference-free metric for evaluating music enhancement algorithms. In: Proceedings of Interspeech (2019)

\bibitem{kong2020panns}
Kong, Q., Cao, Y., Iqbal, T., Wang, Y., Wang, W., Plumbley, M.D.: {PANNs: Large-scale pretrained audio neural networks for audio pattern recognition}. IEEE/ACM Transactions on Audio, Speech, and Language Processing  \textbf{28},  2880--2894 (2020)

\bibitem{li2023mert}
Li, Y., Yuan, R., Zhang, G., Ma, Y., Chen, X., Yin, H., Xiao, C., Lin, C., Ragni, A., Benetos, E., et~al.: Mert: Acoustic music understanding model with large-scale self-supervised training. arXiv preprint arXiv:2306.00107  (2023)

\bibitem{manocha2021cdpam}
Manocha, P., Jin, Z., Zhang, R., Finkelstein, A.: Cdpam: Contrastive learning for perceptual audio similarity. In: ICASSP 2021-2021 IEEE International Conference on Acoustics, Speech and Signal Processing (ICASSP). pp. 196--200. IEEE (2021)

\bibitem{marafioti2020gacela}
Marafioti, A., Majdak, P., Holighaus, N., Perraudin, N.: {GACELA}: A generative adversarial context encoder for long audio inpainting of music. IEEE Journal of Selected Topics in Signal Processing  \textbf{15}(1),  120--131 (2020)

\bibitem{moliner2023solving}
Moliner, E., Lehtinen, J., V{\"a}lim{\"a}ki, V.: Solving audio inverse problems with a diffusion model. In: ICASSP 2023-2023 IEEE International Conference on Acoustics, Speech and Signal Processing (ICASSP). pp.~1--5. IEEE (2023)

\bibitem{musicrecommend2}
Moscato, V., Picariello, A., Sperli, G.: An emotional recommender system for music. IEEE Intelligent Systems  \textbf{36}(5),  57--68 (2020)

\bibitem{nayem2023attention}
Nayem, K.M., Williamson, D.S.: Attention-based speech enhancement using human quality perception modelling. IEEE/ACM Transactions on Audio, Speech, and Language Processing  (2023)

\bibitem{prinz2021end}
Prinz, K., Flexer, A., Widmer, G.: On end-to-end white-box adversarial attacks in music information retrieval. Transactions of the International Society for Music Information Retrieval  \textbf{4}(1),  93--105 (2021)

\bibitem{qin2019imperceptible}
Qin, Y., Carlini, N., Cottrell, G., Goodfellow, I., Raffel, C.: Imperceptible, robust, and targeted adversarial examples for automatic speech recognition. In: International Conference on Machine Learning. pp. 5231--5240. PMLR (2019)

\bibitem{ragano2024nomad}
Ragano, A., Skoglund, J., Hines, A.: Nomad: Unsupervised learning of perceptual embeddings for speech enhancement and non-matching reference audio quality assessment. In: ICASSP 2024-2024 IEEE International Conference on Acoustics, Speech and Signal Processing (ICASSP). pp. 1011--1015. IEEE (2024)

\bibitem{riou2025FiLM}
Riou, A., Gagner{\'e}, A., Hadjeres, G., Lattner, S., Peeters, G.: Zero-shot musical stem retrieval with joint-embedding predictive architectures. In: ICASSP 2025-2025 IEEE International Conference on Acoustics, Speech and Signal Processing (ICASSP). pp.~1--5. IEEE (2025)

\bibitem{saadatpanah2020adversarial}
Saadatpanah, P., Shafahi, A., Goldstein, T.: Adversarial attacks on copyright detection systems. In: International Conference on Machine Learning. pp. 8307--8315. PMLR (2020)

\bibitem{musicrecommned1}
Schedl, M.: Deep learning in music recommendation systems. Frontiers in Applied Mathematics and Statistics  \textbf{5},  457883 (2019)

\bibitem{schonherr2018adversarial}
Sch{\"o}nherr, L., Kohls, K., Zeiler, S., Holz, T., Kolossa, D.: Adversarial attacks against automatic speech recognition systems via psychoacoustic hiding. arXiv preprint arXiv:1808.05665  (2018)

\bibitem{serra2021sesqa}
Serr{\`a}, J., Pons, J., Pascual, S.: Sesqa: semi-supervised learning for speech quality assessment. In: ICASSP 2021-2021 IEEE International Conference on Acoustics, Speech and Signal Processing (ICASSP). pp. 381--385. IEEE (2021)

\bibitem{tzanetakis2002musical}
Tzanetakis, G., Cook, P.: Musical genre classification of audio signals. IEEE Transactions on speech and audio processing  \textbf{10}(5),  293--302 (2002)

\bibitem{wu2022catch}
Wu, X., Rajan, A.: Catch me if you can: Blackbox adversarial attacks on automatic speech recognition using frequency masking. In: 2022 29th Asia-Pacific Software Engineering Conference (APSEC). pp. 169--178. IEEE (2022)

\bibitem{zhang2018unreasonable}
Zhang, R., Isola, P., Efros, A.A., Shechtman, E., Wang, O.: The unreasonable effectiveness of deep features as a perceptual metric. In: Proceedings of the IEEE Conference on Computer Vision and Pattern Recognition. pp. 586--595 (2018)

\end{thebibliography}
\end{document}